ARTICLE ...

# Numerical Simulation of the Cleaning Process of Microchannel by an External Stationary Flow

Lyudmila S. Klimenko[1,2,*] and Boris S. Maryshev[1,3]

[1]Institute of Continuous Media Mechanics, Perm, 614013, Russia
[2]Perm State University, Perm, 614068, Russia
[3]Nizhny Novgorod State University, Nizhny Novgorod, 603022, Russia
[*]Corresponding Author: Lyudmila Klimenko. Email: lyudmilaklimenko@gmail.com

**ABSTRACT**

Clogging of microchannels due to particle adhesion on the walls is a serious problem in many industrial processes. As such, the service life of microfluidic systems is determined by their ability to maintain flow without interruption over a long period of time. Quite often, cleaning of microfluidic and filtration systems by flushing them with a clean liquid flow is used to extend their operating time. A hydrodynamic approach is required for modeling of the cleaning process. In this paper, a mathematical model of microchannel cleaning from fine particles due to interaction with the flow is proposed. The interaction of particles with the flow is described within the Stokes approximation, a random force caused by diffusion is taken into account. The interaction between particles is neglected. The interaction with the channel walls is caused by van der Waals force only. The equations describing the dynamics of particles and flow changes inside the microchannel during their detachment are obtained. The problem is solved numerically in the framework of the random walk model with statistical processing of obtaining data. The flow characteristics and the time of microchannel cleaning depending on the model parameters are obtained.

**Nomenclature**

| | |
|---|---|
| $H$, $L$ | Microchannel width and length |
| $S_d$, $S_u$ | Position of bottom and upper boundary |
| $p$ | Fluid pressure |
| $Q$ | Volume concentration of the settled impurities |
| $V_p$ | Volume occupied by settled impurities |
| $V$ | Volume of the microchannel |
| $\mathbf{n}$ | Unit vector along the normal |

| | | |
|---|---|---|
| $\mathbf{U} = u, w$ | | Velocity of the fluid |
| $A$ | | Hamaker constant |
| $J$ | | Flow rate |
| $T$ | | Fluid temperature |
| $k_B$ | | Boltzmann constant |
| $f_x, f_y$ | | Random variables |
| $a$ | | Particle radius |
| $d$ | | Channel gap |
| $l_m$ | | Minimum distance between the upper and lower boundaries |
| $x, y$ | | Cartesian coordinates |

**Greek symbols**

| | |
|---|---|
| $\varphi$ | Vorticity |
| $\eta$ | Viscosity of the fluid |
| $\rho$ | Density of the fluid |
| $\rho_p$ | Density of the impurity particles |
| $\sigma$ | Viscous stresses |
| $\sigma_T$ | Random stresses due to thermal fluctuations |
| $\sigma_c$ | Critical detachment stresses |

## 1 Introduction

Clogging of microchannels due to particle adhesion on the walls is a serious problem in many processes. These include various filtration systems such as water purification systems, industrial and household filters used, for example, in catalysis or liquid chromatography. Typically, the operating time of a filter element is ultimately limited by the time the filter clogs. Another application that is now becoming more and more popular is various microfluidic systems where particles are either part of the system or are present in the form of dust or other contaminants. The ability to maintain flow for long periods of time dictates the usefulness and lifetime of microfluidic systems, so particle sorption is very undesirable [1]. Quite often, in order to extend the lifetime of both filter and microfluidic systems, a sufficiently powerful flow of working fluid is used. In this case, the fluid is preliminarily cleaned from possible impurities [2].

There are several physical mechanisms that lead to clogging of a microchannel. The simplest case of clogging is the "mechanical" blockage, when particles get into the channel, the characteristic gap of which is smaller than their own size [3, 4]. This approach has experimental confirmation and describes the transport of sufficiently large particles. However, it is known that clogging is observed even when a suspension containing finely dispersed particles flows through the channel.

Blockage by small particles is often explained by the formation of aggregates resulting from particle-particle interaction, and then a "new" larger particle can lead to mechanical blockage [5]. This behavior was observed experimentally in different regions of the channel (near the wall and in the liquid volume) and for particles of different sizes [6, 7]. Later it was shown numerically [8], [9]. The authors explain clogging by mutual aggregation of particles, but sorption of particles on the wall and effective narrowing of real channels are not taken into account. However, in some cases, it is the narrowing of channels due to sorption on the wall that turns out to be the determining factor leading to channel clogging, which was shown in [10].

There are also many works devoted to studies of microfluidic devices. Thus it was shown experimentally in [11] that collective effects, such as the formation of aggregates, do not play an important role in the clogging of the device. Numerically [12] using an adhesion/collision model (JKR) that accounts for aggregate formation as well as particle deposition on the channel walls, this was confirmed. However, the influence of the channel shape on the flow was not considered, such influence was taken into account in [10].

The paper [10] is devoted to the study of clogging of an initially clean channel by a fine impurity, taking into account the mutual influence of the shape of the walls (which changes when particles are deposited on them) on the fluid flow formed in the channel gap under the action of an external pressure drop. In this work, the disruption of particles from the walls is provided by the action of thermal fluctuations (affecting the strength of the bond between the particle and the wall) and viscous forces arising from the flow. The same approach is developed in the present work to study the possibility of cleaning a clogged channel by the flow of pure liquid.

Thus, there are many experimental and numerical studies of microchannel blockage. At the same time, the inverse problem of channel cleaning from finely dispersed impurities has not been practically investigated. Perhaps, its solution is considered to be a trivial generalization of the problem of channel blockage, but, as the present study has shown, cleaning is not always possible and represents an additional independent study.

## 3 Problem Statement

We consider a rectangular microchannel $H \times L$ filled with viscous fluid. At the initial time moment the channel is clogged with fine impurities, the impurity particles are held on the walls and among themselves by van der Waals forces. To clean the microchannel, a pure fluid is pumped horizontally by the imposed constant pressure drop between the channel inlet and outlet $\Delta P = P_1 - P_2 > 0$. Since, at the initial moment the walls of the channel are uneven, the viscous stresses arising on the channel walls may be sufficient for particle detachment by the flow. Then the detachment particle is transported by the fluid, driving by random force, caused by diffusion, while interacting with the channel walls. The form of walls is changed during the process of particles detachment and attachment. The forms of bottom and upper walls in each moment of time are described by the functions $S_d$ and $S_u$ respectively. The scheme of the problem under consideration is presented in Fig. 1.

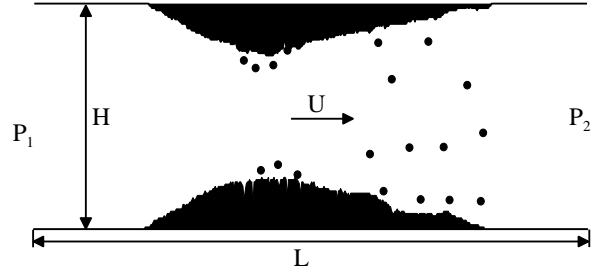

**Figure 1**: Sketch of the problem

To calculate the flow in a channel with a complex wall shape formed by particles, the model developed in [10] for laminar flow of a viscous fluid at small Reynolds numbers is used. In dimensionless units the problem describing a stationary flow in terms of vorticity ($\varphi$) and pressure ($p$) can be written in the following form

$$\frac{\partial^2 p}{\partial x^2}+\frac{\partial^2 p}{\partial y^2}=0, \quad \frac{\partial^2 \varphi}{\partial x^2}+\frac{\partial^2 \varphi}{\partial y^2}=0,$$

$$p\big|_{x=0}=1, \quad p\big|_{x=L}=0, \quad \frac{\partial p}{\partial \vec{n}}\bigg|_{y=S_l,S_u}=0, \quad (1)$$

$$\frac{\partial \varphi}{\partial x}\bigg|_{x=0,L}=0, \frac{\partial \varphi}{\partial \mathbf{n}}\bigg|_{y=S_u}=\frac{\partial p}{\partial \tau}\bigg|_{y=S_u}, \quad \frac{\partial \varphi}{\partial \mathbf{n}}\bigg|_{y=S_l}=-\frac{\partial p}{\partial \tau}\bigg|_{y=S_l},$$

where $S_l(x), S_u(x)$ are function of x coordinate, describing the shape of bottom and top channel walls, $\mathbf{n},\tau$ are unit vectors along the normal and along the channel walls, respectively.

Problem (1) is written in dimensionless form. We use the following units for pressure, distance, velocity and time as $P_1-P_2$, $H$, $H\,P_1-P_2\,H\eta^{-1}$, $8L^2\eta H^{-2}\,P_1-P_2^{\,-1}$, where $\eta,\rho$ are dynamic viscosity and density of the fluid.

The velocity of the fluid flow $\mathbf{U}=u,w$ can be calculated as a solution of

$$\frac{\partial^2 u}{\partial x^2}+\frac{\partial^2 u}{\partial y^2}=\frac{\partial p}{\partial x}, \quad \frac{\partial^2 w}{\partial x^2}+\frac{\partial^2 w}{\partial y^2}=\frac{\partial p}{\partial y}, \quad \varphi=\frac{\partial w}{\partial x}-\frac{\partial u}{\partial y},$$

$$\frac{\partial u}{\partial y}\bigg|_{x=0,L}=-\varphi\big|_{x=0,L}, u\big|_{y=S_l,S_u}=0,$$

$$\frac{\partial u}{\partial y}\bigg|_{x=0,L}=-\varphi\big|_{x=0,L}, \quad u\big|_{y=S_l,S_u}=0, \quad (2)$$

$$w\big|_{x=0,L}=0, \quad w\big|_{y=S_l,S_u}=0,$$

It allows to calculate the shear stress. If the shear stress value greater then van der Waals force between particle and channel walls, the particle will detach and drift with the flow governed by the following equation system

$$\begin{aligned}x=tu\,x,y,t\,+af_x\,t\,,\\ y=tw\,x,y,t\,+af_y\,t\,,\end{aligned} \quad (3)$$

where $f_x, f_y$ is vector of random variables distributed according to the normal law (with unit variance and zero mean). We assume that impurity particles are solid, uncharged and identical with radius $a$, which obeys the following restriction: $a \ll a_c = 18\pi\eta^2 H^2 / \rho_p k_B T$, where $\rho_p$ is the particle density, $k_B$ is the Boltzmann constant, $T$ is the temperature of the fluid. Here the dominant transport mechanism is particle transport by fluid flow, as it was shown in [10].

Since channel cleaning is considered, the particle can start drifting in the flow only after detachment from the wall, the detachment conditions according to [10] can be written in the following form

$$\sigma|_{y=S_l} = \frac{\partial u}{\partial \mathbf{n}} + \sigma_T f_l > \sigma_c,$$

$$\sigma|_{y=S_u} = -\frac{\partial u}{\partial \mathbf{n}} + \sigma_T f_u > \sigma_c,$$
(4)

where $f_u, f_l$ are independent random variables distributed by normal law, $\sigma_T = 3k_B T |P_1 - P_2|^{-1} / \sqrt{2a^5 H}$ is a dimensionless value of random stresses due to thermal fluctuations, $\sigma_c = 2A |P_1 - P_2|^{-1} / 27a^3$ is dimensionless value of critical detachment stresses (caused by the van der Waals interaction), $A$ is the Hamaker constant [13]. For most existing materials, the values of the Hamaker constant vary in the range of $A = 10^{-19} \div 10^{-22}$ J. Thus dimensionless values of critical stresses can vary in a wide range, let us give as an example possible estimates of parameters of the problem. For a microchannel with a width of $H \sim 10\,\mu m$, whose walls are made of silicon dioxide ($SiO_2$), and ferric oxide ($Fe_2O_3$) particles with $a \approx 10^{-7} m \ll a_c$ for small pressure drop as $P_1 - P_2 \approx 10\,Pa$, the following estimates for dimensionless parameters can be obtained: for water $\mathrm{Re} \approx 0.02$, $\sigma_T \approx 0.09$ for $T = 300K$, $\sigma_c \approx 0.31$, with the Hamaker constant for the given materials $A \approx 4.2 \cdot 10^{-20}$ J.

## 4 Numerical method

In the numerical scheme the channel with aspect ratio $L/H = 5$ is used. The problem is solved on a uniform grid with 100 grid nodes per unit length, so the channel is covered by a grid of nodes. The particles are assumed to be identical, their size , i.e. a particle occupies one node of the grid. The self-adjoint problem (1) is solved numerically with finite difference discretization of the second order of accuracy. It is assumed that particle detachment is slow enough, for the obtained stationary flow, conditions (4) are checked for each node along the current wall position. Then, in those nodes where the condition is satisfied, the particle detachment from the wall is carried out. In this case, the wall coordinate is shifted by one node, and the particle transport with the flow is considered taking into account random jumps caused by thermal fluctuations (diffusion).

In this case, the particle can again come close to the wall (if at some moment it makes a jump towards the wall), in which case it adheres to the wall in a new place, the position of the wall at this node is shifted by one node, and the calculation of the particle motion is suspended. However, the particle may reach the exit from the channel without approaching the wall, then this particle is considered to have flown out of the channel and the calculation is also suspended. This procedure is repeated for all nodes along the wall where conditions (4) have been satisfied. As a result, the shape of the channel walls is slightly changed, after which a new steady-state flow calculation with slightly changed channel geometry is performed. In this case, the previous distribution of vorticity and pressure fields is taken as an initial approximation, since the geometry changes only slightly, and the problem (1) is solved by the iterative method, the calculation converges rather quickly (usually 3-5 iterations are required). In this calculation scheme, it is assumed that the particles break off rarely enough and do not affect each other in the flow. It was shown in [10] that such an approximation is justified for particles of size

$$a \ll a_c = \frac{18\pi\eta^2 H^2}{\rho_p k_B T}$$

The time of the process at each step is defined as the maximum of all times necessary for particles to "travel" from the moment of detachment to exit from the channel or to stick in another place when solving equations of motion (3). In this case, as shown in [10], the most optimal time step (for solving equations (3)) is given by the relation

$$\delta t = \frac{3\pi\eta a^3}{2k_B T}$$

Since the flow calculation is performed by iterative method, it is necessary to have an adequate initial approximation for convergence. If the initial distribution of particles in the "polluted" channel is set randomly, it is not possible to obtain the initial distribution of pressure and swirl, there will be divergence. Therefore, to "prepare" the initial distribution of particles, we plug the channel using the scheme described above, but without using the conditions (4), i.e., assuming that the particles do not bounce off the walls. At each step, at the inlet to the channel, the position of the particle is randomly set, after which its drift in the flow is calculated, until the moment of adhesion to the wall or exit from the channel. In case of sticking, the flow in the channel is recalculated and a new particle is started. In this case, at the first step, the channel is completely clear and the flow in it is Poiseuille, i.e., the following relations can be chosen as an initial approximation for pressure and vorticity:

$$p = 1 - \frac{x}{L}, \quad \varphi = \frac{1}{2} - \frac{y}{L}. \tag{5}$$

The calculation of channel "clogging" ends when the critical gap is reached, the distance between the upper and lower walls at which the fluid flow through the channel drops by a factor of 10 (when it is no longer possible to use the approximation of non-interacting particles in the flow). The validity and accuracy of this approach has been demonstrated in [10].

## 4 Numerical Results

As a result of calculations, the dynamics of channel cleaning and evolution of flow field have been calculated. The most typical picture of cleaning process is shown in Figure 2. Depending on problem parameters, the initially clogged microchannel quickly becomes clean everywhere except for the largest particle deposits (see Figure 2, b). Complete cleaning can be achieved, as will be shown later, in some parameter range, otherwise the channel remains clogged.

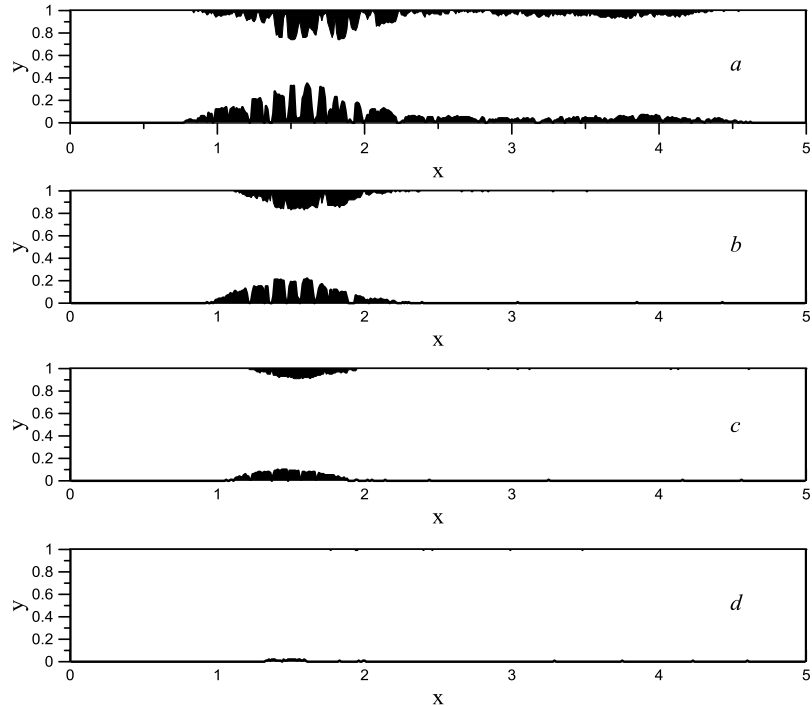

**Figure 2**: The position of the walls of the channel contaminated with impurity, the area occupied by impurity particles is shaded in black. Fragments of figure a-d correspond to different moments of time: a – $t=10$, b – $t=50$, c – $t=50$, d – $t=120$. Complete cleaning of the channel was achieved at the moment of time $t=140$.

Since the cleaning process is predominantly random, important information can be extracted from the integral parameters. For this purpose we performed 21 independent realizations for each value of problem parameters. The resulting value of integral parameter is the average of these realizations.

Figure 3 shows the dependence of the volume concentration of the settled impurities ($Q=V_p/V$, where $V_p$ is the volume occupied by the settled impurity, $V$ is the volume of the whole channel space) on time for different values of the parameter $\sigma_T$.

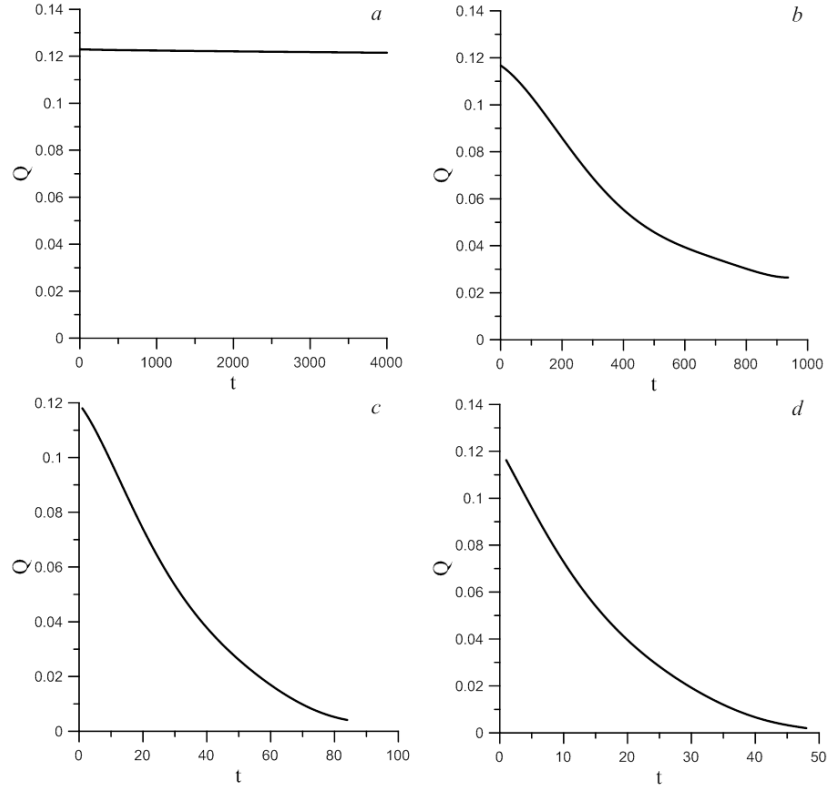

**Figure 3**: Dependences of the volume concentration of deposited impurity ($Q$) on time for different values of the parameter $\sigma_T$: a - $\sigma_T = 0.05$, b - $\sigma_T = 0.08$, c - $\sigma_T = 0.11$, d - $\sigma_T = 0.15$

From Figure 3 it is clear that with increasing of temperature or $\sigma_T$ the channel cleaning is strongly accelerated, so at $\sigma_T = 0.11$, the cleaning time is reduced by a factor of 10 (compared to $\sigma_T = 0.08$ Fig.3b and Fig.3c )

For smaller values of $\sigma_T$ the microchannel cannot be cleaned completely, the concentration value remains practically unchanged for a long time. Such dynamics can be clearly seen on the time dependence of the channel gap ($d = l_m / H$, where $l_m$ is the minimum distance between the upper and lower boundaries) is shown in Figure 4. At high values of $\sigma_T$ the gap value becomes equal to 1 rather quickly (fig.4 d), which indicates complete cleaning of the channel. At the threshold value $\sigma_T = 0.08$ a significant slowdown of the cleaning process is observed. Thus, here the contribution of viscous stresses is rather small, which leads to actual stopping of the cleaning process. Indeed, for $\sigma_T = 0.05$ the gap in general has no clear tendency neither to increase nor to decrease, there are little noticeable chaotic oscillations caused by thermal fluctuations, the channel cleaning does not occur.

Another important characteristic of flow through a microchannel is the fluid flow rate at the channel outlet, which can be determined as follows $J = \int_0^H u(x = L, y) dy$.

The time dependence of the flux is presented in Figure 5. All lines of Figure 5 show a tendency to flow rate growth as thermal fluctuations $\sigma_T$ increase.

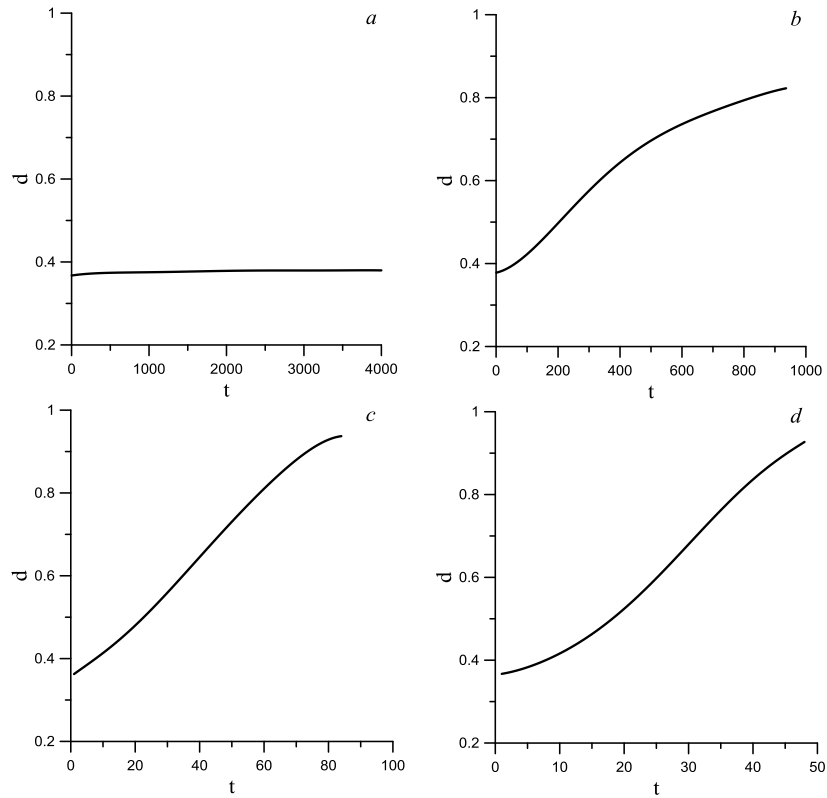

**Figure 4**: Dependences of the dimensionless channel gap ($d$) on time for different values of the parameter $\sigma_T$: a - $\sigma_T = 0.05$, b - $\sigma_T = 0.08$, c - $\sigma_T = 0.11$, d - $\sigma_T = 0.15$

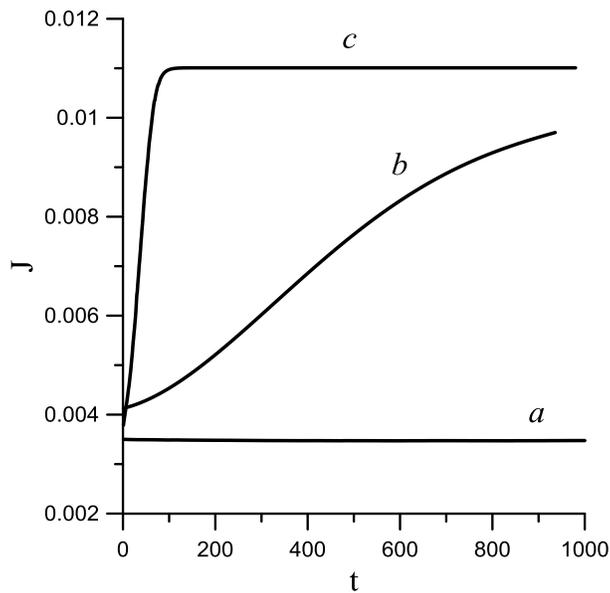

**Figure 5:** Time dependences of the flow rate through the channel ($J$) for different values of the parameter $\sigma_T$: **a -** $\sigma_T = 0.05$, **b -** $\sigma_T = 0.08$, **c -** $\sigma_T = 0.11$

Dependences of the flow rate on the volume concentration of settled impurity are presented in Fig. 6. It can be seen that all characteristics have a well-defined linear part at small values of concentration, which is quite consistent with Darcy's law (as shown in [10]) for porous media or the Batchelor's corelation for microchannels (see [10] and [14]).

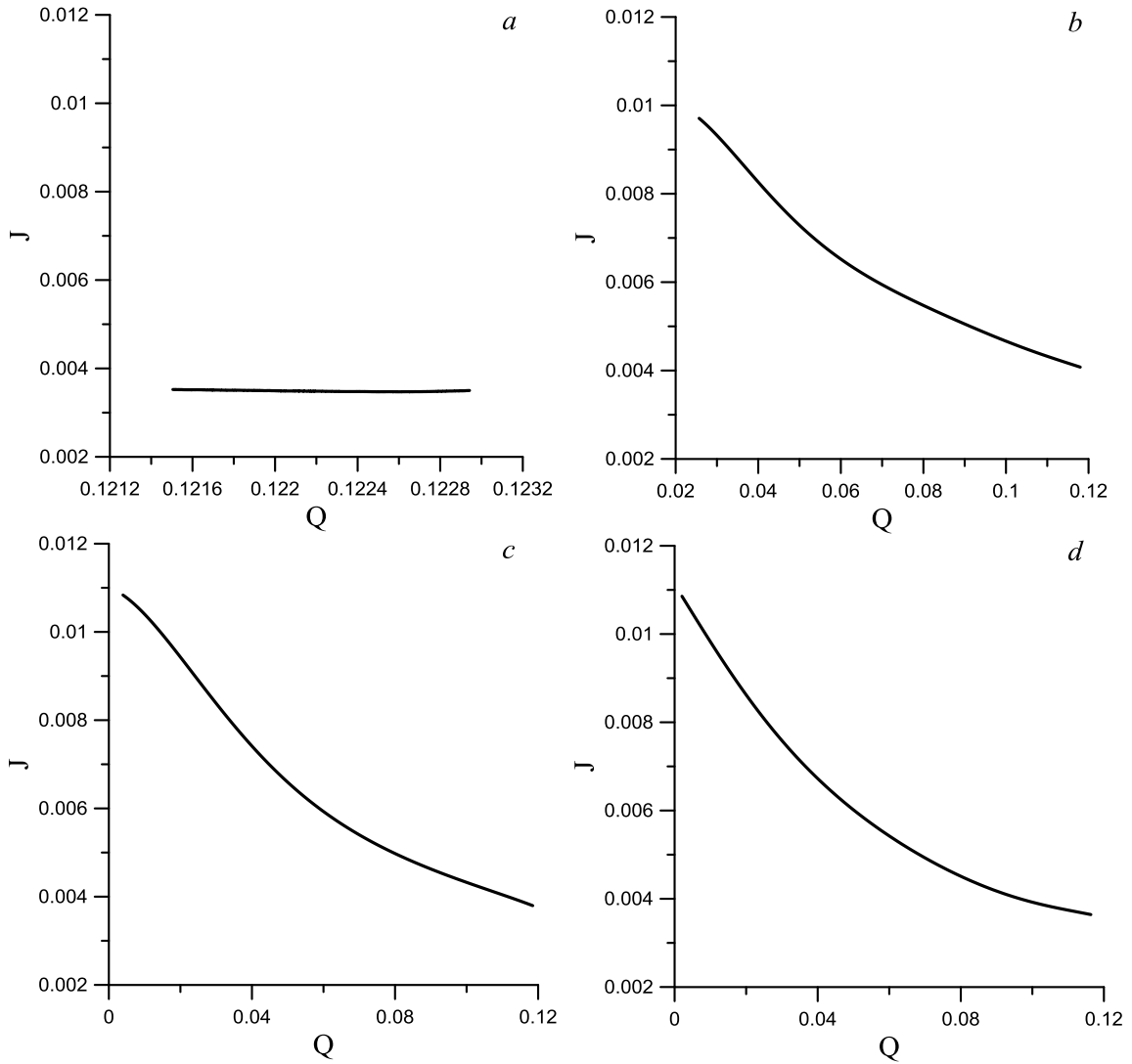

**Figure 6**: Dependences of the flow rate through the channel ($J$) on the impurity volume concentration ($Q$) for different values of the parameter $\sigma_T$: a - $\sigma_T = 0.05$, b - $\sigma_T = 0.08$, c - $\sigma_T = 0.11$, d - $\sigma_T = 0.15$

## 5 Conclusion

The process of cleaning of a clogged microchannel by external imposed steady flow has been studied. The assumptions of laminar flow and weak concentration of suspended particles have been applied. It is shown that very slow channel cleaning occurs under any pumping conditions and its cause is random thermal fluctuations, when a particle can randomly leave its place on the wall. However, when the average value of viscous stresses on the wall becomes larger than the bond value, the cleaning process can be accelerated by tens or even hundreds of times. It is shown that when describing the process, the assumptions of laminar flow remain valid, and Darcy's law for a porous medium or the equivalent Betschelor's relation for microchannels are fulfilled with good accuracy.

**Funding Statement:** The authors acknowledge the financial support from RSF (Grant no. 23-12-00180)

**Author Contributions:** Study conception and mathematical model: B.M.; numerical study: L.K.; analysis and interpretation of results: B.M., L.K.; draft manuscript preparation: L.K. All authors reviewed the results and approved the final version of the manuscript.

**Availability of Data and Materials:** The data that support the findings of this study are available from the corresponding author upon reasonable request.

**Conflicts of Interest:** The authors declare that they have no conflicts of interest to report regarding the present study.